\documentclass[prl,twocolumn,superscriptaddress]{revtex4}

\usepackage{comment}

\usepackage{amsmath,amssymb,amsfonts}
\usepackage{bm}
\usepackage{graphicx}

\begin{document}

\title{Coulomb interaction effect in tilted Weyl fermion in two dimensions}

\author{Hiroki Isobe%
\footnote{Present address: Massachusetts Institute of Technology, Cambridge, Massachusetts, 02139, USA.} 
}
\affiliation{Department of Applied Physics, University of Tokyo, Bunkyo, Tokyo 113-8656, Japan}

\author{Naoto Nagaosa}

\affiliation{RIKEN Center for Emergence Matter Science (CEMS), Wako, Saitama 351-0198, Japan}
\affiliation{Department of Applied Physics, University of Tokyo, Bunkyo, Tokyo 113-8656, Japan}

\date{\today}

\begin{abstract}
Weyl fermions with tilted linear dispersions characterized 
by several different velocities appear in some systems including the quasi-two-dimensional organic 
semiconductor $\alpha$-(BEDT-TTF)$_2$I$_3$ and three-dimensional WTe$_2$. 
The Coulomb interaction between electrons modifies the 
velocities in an essential way in the low-energy limit, where the logarithmic
corrections dominate. 
Taking into account the coupling to both the transverse and longitudinal electromagnetic fields,
we derive the renormalization group equations for the velocities 
of the tilted Weyl fermions in two dimensions, and found that
they increase as the energy decreases and eventually 
hit the speed of light $c$ to result in the Cherenkov radiation.
Especially, the system restores the isotropic Weyl cone even when the bare Weyl cone is strongly tilted and the velocity of electrons becomes negative in certain directions.  
\end{abstract}


\maketitle

Weyl fermions (WFs) in solids are the subject of intensive 
interest recently from the viewpoint of their topological aspects such
as the chiral anomaly~\cite{ca1,ca2,ca3,ca4} as well as
the associated novel transport properties~\cite{tp1,tp2,tp3,tp4,mr1,mr2,mr3,mr4}. 
They act as the (anti-)monopoles 
of the Berry curvature in momentum space, and hence are characterized 
by the winding number or magnetic charge.

Since the density of states (DOS) vanishes at the zero energy, the Coulomb interaction 
is not screened enough, and its effect on Weyl fermions has been studied 
theoretically with the focus on two-dimensional (2D) graphene~\cite{graphene1,graphene2,graphene3} and three-dimensional (3D) Weyl 
semimetals~\cite{3d_int}. These studies have revealed that the velocity of electrons 
is enhanced as the energy decreases in sharp contrast to the conventional case
where the mass is enhanced and velocity is reduced by the electron correlation~\cite{nozieres}. 
The logarithmic enhancement of velocity is experimentally observed in graphene by Shubnikov--de Haas oscillations, which was explained by a renormalization group (RG) analysis of the Coulomb interaction for isotropic WFs~\cite{graphene_exp}. 

The logarithmic divergence of the velocity due to the Coulomb interaction, however,
is unphysical when it exceeds the speed of light $c$. This problem is resolved
when one takes into account the coupling between the electrons and the 
transverse part of the electromagnetic field, i.e., photon. In this case, the
electron velocity $v$ approaches to $c$ in the low-energy limit~\cite{gonzalez,3d}.
  
\begin{figure}
\centering
\includegraphics[width=\hsize]{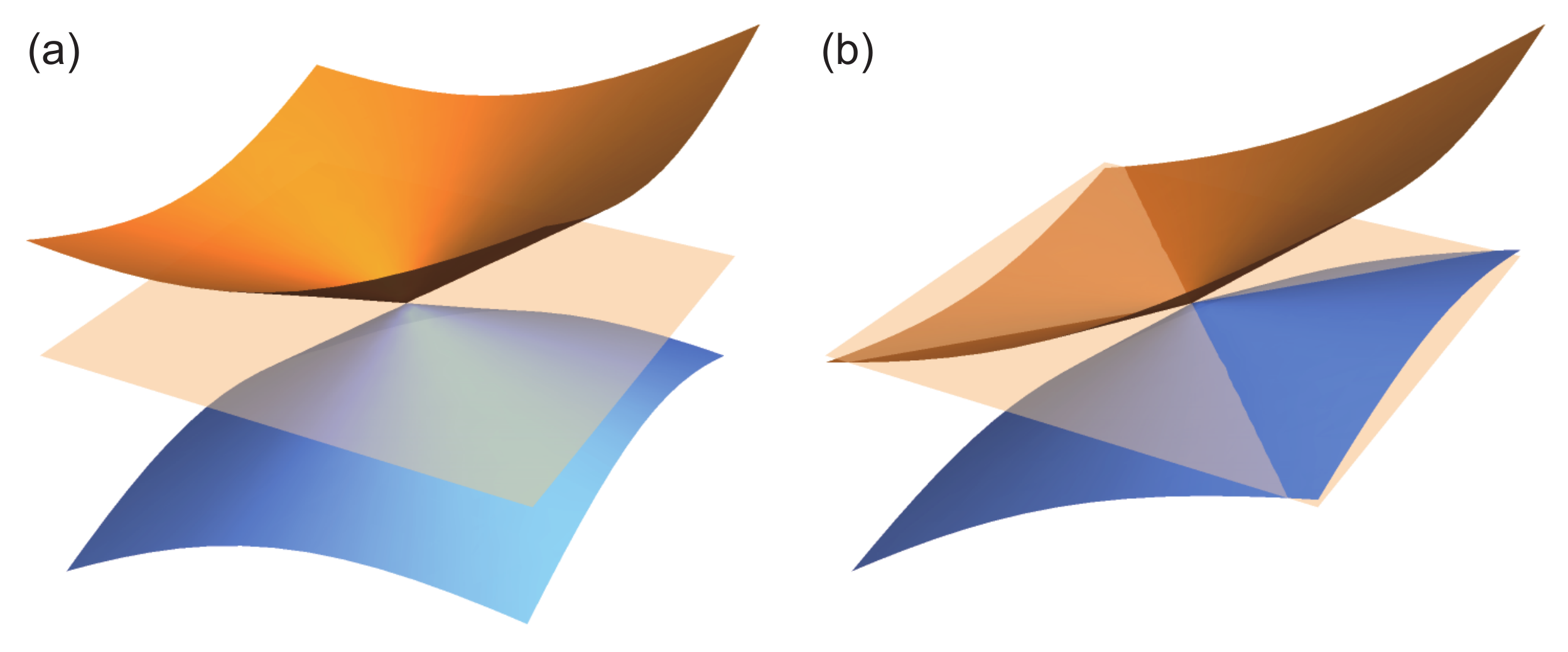}
\caption{
Two types of the energy dispersions of the tilted WFs. 
(a) Type I, with $v_i > w_i$ for all $i$. The conduction and valence 
bands touch at a single point, $k=0$. 
(b) Type II. There are electron and hole pockets, which are bounded by lines and touch at $k=0$.  
In this figure, we set $v_i < w_i$ for a single $i$.
}
\label{fig:1}
\end{figure}

An interesting generalization of WFs is to consider asymmetric velocity parameters, i.e.,
\begin{equation}
H = \sum_{i} (v_i \sigma^i + w_i)  k_i ,
\label{eq:Hamil}
\end{equation}
where $\sigma^i$ is a Pauli matrix.
We set $\hbar =1$ throughout the paper. 
$i$ runs from 1 to 2 (3) for 2D (3D) systems.
$v_i$'s can take different values, and 
$w_i$'s describe the tilt of the linear energy dispersion depending on the direction. 
This is named a ``tilted'' Weyl fermion.
With this Hamiltonian, there are two types of tilted WFs; type I possesses only positive velocities 
[Fig.~\ref{fig:1}(a)], and type II has a negative velocity in some direction 
[Fig.~\ref{fig:1}(b)]~\cite{bernevig}. 
Equation~\eqref{eq:Hamil} is not unrealistic. Actually, 
organic compound $\alpha$-(BEDT-TTF)$_2$I$_3$ is a
quasi-2D conductor which supports the 2D 
WFs~\cite{bender,kino,tajima2009,kobayashi2009}.
The crystal symmetry of this material is low enough,
which results in the two tilted WFs at $\bm{k}_0$ and 
$-\bm{k}_0$. Especially, it is believed that the electron correlation is 
appreciable here because of the proximity to the charge ordering in the 
phase diagram. Actually, $^{13}$C-NMR experiment 
under hydrostatic pressure has been analyzed successfully 
by the renormalization effect of the velocities 
due to the Coulomb interaction~\cite{jpsj}. 
Another example is WTe$_2$, which shows novel magnetotransport 
properties~\cite{cava}. It is proposed that the electronic states of 
this 3D material are described by Eq.~(\ref{eq:Hamil}) 
even with the opposite signs of velocities~\cite{bernevig}. 

In this Letter, we study the effect of coupling between electrons and
longitudinal (Coulomb) and transverse electromagnetic fields on 
the velocities in 2D tilted WFs.
Especially there are two issues. 
One is how the speed of light $c$ and the tilt of the 2D WF enter into the renormalization 
of electron's velocities. 
The other is the change of the Fermi surface due to the interactions. Namely, the Fermi surface is a point for the same sign of the velocities in the two directions, while it consists of two lines in the case of opposite signs.  
It will be shown below that for both type I and type II WFs, the electron's velocities are renormalized to approach the speed of light $c$ and hence the system recovers the Lorentz symmetry in the low-energy limit. 
Also, the interaction brings the change in the topology of the Fermi surface of type II WFs,
leading to the third class. In this case the electron's velocities hit the speed of light $c$, which results in the Cherenkov radiation.

We consider the action that describes a tilted WF in a (2+1)D plane placed in the (3+1)D space. 
We assume that the WF is confined on the $xy$ plane and that the Weyl cone is tilted along the 
$x$ direction, for simplicity. 
Then the action $S$ becomes 
\begin{equation}
\label{eq:action}
S = \int d^3 x \mathcal{L}_e + \int d^4 x \mathcal{L}_\gamma, 
\end{equation}
where $\mathcal{L}_e$ and $\mathcal{L}_\gamma$ are given by 
\begin{gather}
\mathcal{L}_e = \bar{\psi}(x) i[ \gamma^0 (D_0 + wD_1) + v_x \gamma^1 D_1 
+ v_y \gamma^2 D_2] \psi (x), \\
\mathcal{L}_\gamma = \frac{1}{2} \left( \varepsilon \bm{E}^2 - \frac{1}{\mu} 
\bm{B}^2 \right). 
\end{gather}
$w$ determines the tilt of the Weyl cone, whose velocities are described by $v_x \pm w$ and $v_y$. 
$D_{\mu}$ is the gauge covariant derivative, given by $D_{\mu} = \partial_{\mu} + ieA_{\mu}$. 
$\psi (x)$ and $A_\mu(x)$ correspond to the electron field and the vector potential 
for the electromagnetic field. 
We work in the Minkowski space with the metric tensor $g_{\mu\nu} = \text{diag} (+1, -1, -1, -1)$. 
The gamma matrices satisfy the anticommutation relation 
$\{ \gamma^\mu, \gamma^\nu\} = 2g^{\mu\nu}$. 
The electron propagator $S_0(k)$ is given by
\begin{equation}
S_0 (k) = \frac{i}{\gamma^0(k_0 + wk_1) + v_x \gamma^1 k_1 + v_y \gamma^2 k_2 + i0^+}. 
\end{equation}
The noninteracting vertex $\Gamma_0^{\mu}$ is 
\begin{equation}
\Gamma_0^{\mu} = -iel^{\mu}_{\nu} \gamma^{\nu} \quad (\mu, \nu= 0,1,2), 
\end{equation}
where the matrix $l^{\mu}_{\nu}$ is defined by
\begin{equation}
l^{\mu}_{\nu} =
\begin{pmatrix}
1 & 0 & 0 \\
w/c & v_x/c & 0 \\
0 & 0 & v_y/c
\end{pmatrix}. 
\end{equation}
The electric and magnetic fields $\bm{E}$ and $\bm{B}$ are represented by using 
the gauge field $A_\mu$ as 
\begin{gather}
\bm{E} = -\frac{1}{c} \frac{\partial \bm{A}}{\partial t} - \nabla A_0, \quad
\bm{B} = \nabla \times \bm{A}. 
\end{gather}
The speed of light $c$ in a material is determined by the relative permittivity $\varepsilon$ and the relative permeability $\mu$ as $c=c_0/\sqrt{\varepsilon\mu}$, where $c_0$ is the speed of light in vacuum. 
In the following analysis, we employ the Feynman gauge, and thus the electromagnetic 
field propagator in (3+1)D is given by 
\begin{align}
D_0^{\mu\nu} (q) = \frac{-ic^2 g^{\mu\nu}}{\varepsilon (q^2 + i0^+)} ,
\end{align}
with $q^2 = q_0^2 -c^2 \bm{q}^2$. 
When we focus on the (2+1)D plane where electrons are confined, $D_0^{\mu\nu}(q)$ 
is reduced to be 
\begin{align}
\tilde{D}_0^{\mu \nu} (q) = \int \frac{dq^3}{2\pi} D_0^{\mu\nu} (q) 
= \frac{ic g^{\mu \nu}}{2\varepsilon\sqrt{-\tilde{q}^2}},  
\end{align}
with $\tilde{q}^2 = q_0^2 -c^2(q_1^2+q_2^2)$. 

We analyze the effect of the electron-electron interaction mediated by the electromagnetic 
field to one-loop order. 
Here we include both the transverse and longitudinal parts of the electromagnetic field. 
We note that the polarization at one loop is not divergent in (2+1)D 
as well as that of graphene~\cite{gonzalez}, and thus the speed of light $c$ is not renormalized. 
We set $c=1$ for simplicity in the following analysis. 
Also the Ward-Takahashi identity guarantees the relation between the self-energy and the 
vertex correction. 
The self-energy at one-loop order is given by
\begin{align}
\label{eq:self}
-i \Sigma (p) = (-ie)^2 \int \frac{d^d k}{(2\pi)^d} l^{\mu\nu} 
\gamma_\nu S_0(k) l^{\lambda\sigma} \gamma_\sigma \tilde{D}_{0,\mu\lambda} (p-k). 
\end{align}
To regularize the divergent integral, we employ the dimensional regularization; the dimension 
of spacetime is shifted as $d=3-\epsilon$. 

From the self-energy in eq.~\eqref{eq:self}, we obtain the coupled RG equations for $v_x$, $v_y$, and $w$~\cite{SM}: 
\begin{widetext}
\begin{align}
\label{eq:rg_vx}
\kappa \frac{dv_x}{d\kappa} &= \beta_{v_x} (v_x, v_y, w) =
-\frac{g^2}{2\pi} v_x 
\left[ 2(1-v_x^2) F_1^0 -(1-v_x^2) (1-w^2-v_x^2-v_y^2) F_1^1 \right], \\
\label{eq:rg_vy}
\kappa \frac{dv_y}{d\kappa} &= \beta_{v_y} (v_x, v_y, w) =
-\frac{g^2}{2\pi} v_y
\left[ 2 (1-w^2-v_y^2) F_2^0 -g_1 F_2^1 +g_2 F_2^2 \right], \\
\label{eq:rg_w}
\kappa \frac{dw}{d\kappa} &= \beta_w (v_x, v_y, w) =
-\frac{g^2}{2\pi} w 
\left[ -2v_x^2 F_1^0 + v_x^2 (1+ w^2-v_x^2-v_y^2) F_1^1 \right], 
\end{align}
\end{widetext}
where $\kappa$ is the renormalization scale, and $g^2=e^2/(4\pi\varepsilon) \approx (1/137)/\varepsilon$ 
is a dimensionless constant, with the definition $c=1$. 
$F_{1,2}^n$ and $g_{1,2}$ are functions which are defined as follows:
\begin{gather*}
F_1^n (v_x, v_y, w) = \int_0^1 dx \frac{x^n \sqrt{1-x}}{f_1^{3/2} (x; v_x, v_y) f_2^{1/2} 
(x; v_y)}, \\
F_2^n (v_x, v_y, w) = \int_0^1 dx \frac{x^n \sqrt{1-x}}{f_1^{3/2} (x; v_x, v_y) f_2^{3/2} 
(x; v_y)},
\end{gather*}
with 
\begin{gather*}
f_1 (x; v_x, w) = 1-x(1-v_x^2) - x(1-x) w^2, \\
f_2 (x; v_y) = 1-x(1-v_y^2),
\end{gather*}
and 
\begin{align*}
g_1 (v_x, v_y, w) 
=& (1-w^2+v_x^2-v_y^2)(1-v_x^2+w^2) \notag \\
&+ (1-w^2-v_x^2-v_y^2)(1-v_y^2) \notag \\
&+ (1-v_x^2)(1-w^2-v_x^2-v_y^2) -2w^2 v_x^2, \\
g_2 (v_x, v_y, w) 
=& (1-v_x^2) (1-v_y^2) (1-w^2-v_x^2-v_y^2) \notag \\
&+w^2 (1-w^2+v_x^2-v_y^2) -2 w^2 v_x^2 (1-v_y^2). 
\end{align*}

\begin{figure}
\centering
\includegraphics[width=\hsize]{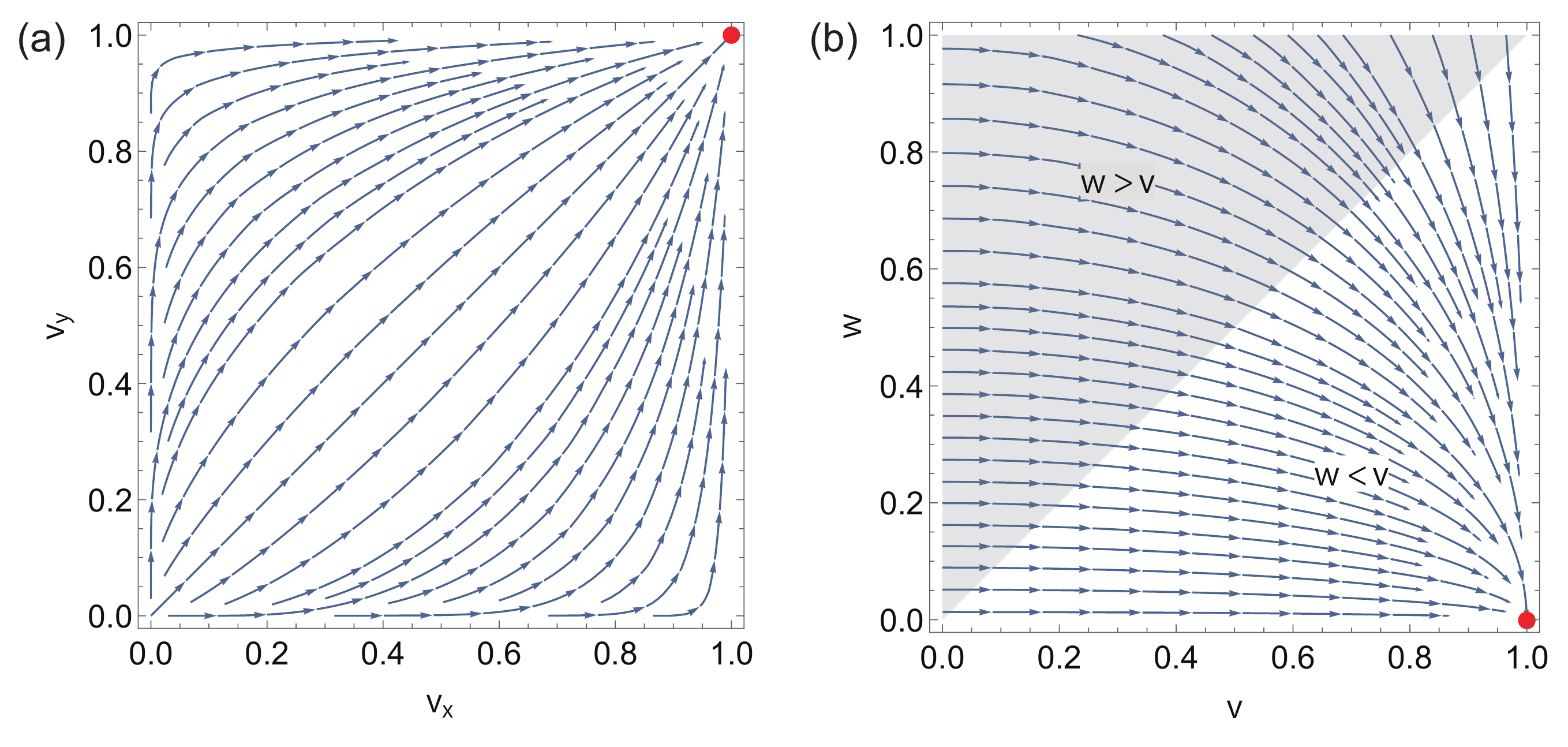}
\caption{
RG flows of the velocity parameters. 
(a) Flow of $v_x$ and $v_y$ without tilting $w=0$ and
(b) flow of $v$ and $w$ with $v\equiv v_x =v_y$. 
The shaded region corresponds to type II where $w>v_x$. 
From those two figures, we can identify that the red point at $v_x=v_y=1$ and $w=0$ 
is the only stable fixed point. 
}
\label{fig:2}
\end{figure}

The RG equations in eqs.~\eqref{eq:rg_vx}--\eqref{eq:rg_w} can be solved analytically for some special cases~\cite{SM}. 
It suffices to consider $v_x, v_y, w \geq 0$, and the flow of the velocity parameters are shown 
in Fig.~\ref{fig:2}. 
There is an infrared stable fixed point at $v_x=v_y=1$ and $w=0$. Therefore, the tilt of the Weyl cone vanishes in the low-energy limit and the energy dispersion 
becomes isotropic with the velocity of electrons being the same as that of light in the material. 
Remarkably, this result applies to both type I and type II. 
It has been known that the tilt is not renormalized if we take into account only
the instantaneous Coulomb interaction~\cite{jpsj}. 
The renormalization of the tilt $w$ arises from the relativistic effect, i.e., the coupling 
of the electron field to the transverse electromagnetic field. 
Thus the renormalization of $w$ is stronger for large velocities, 
as we can see from Fig.~\ref{fig:2}(b).

\begin{figure}
\centering
\includegraphics[width=\hsize]{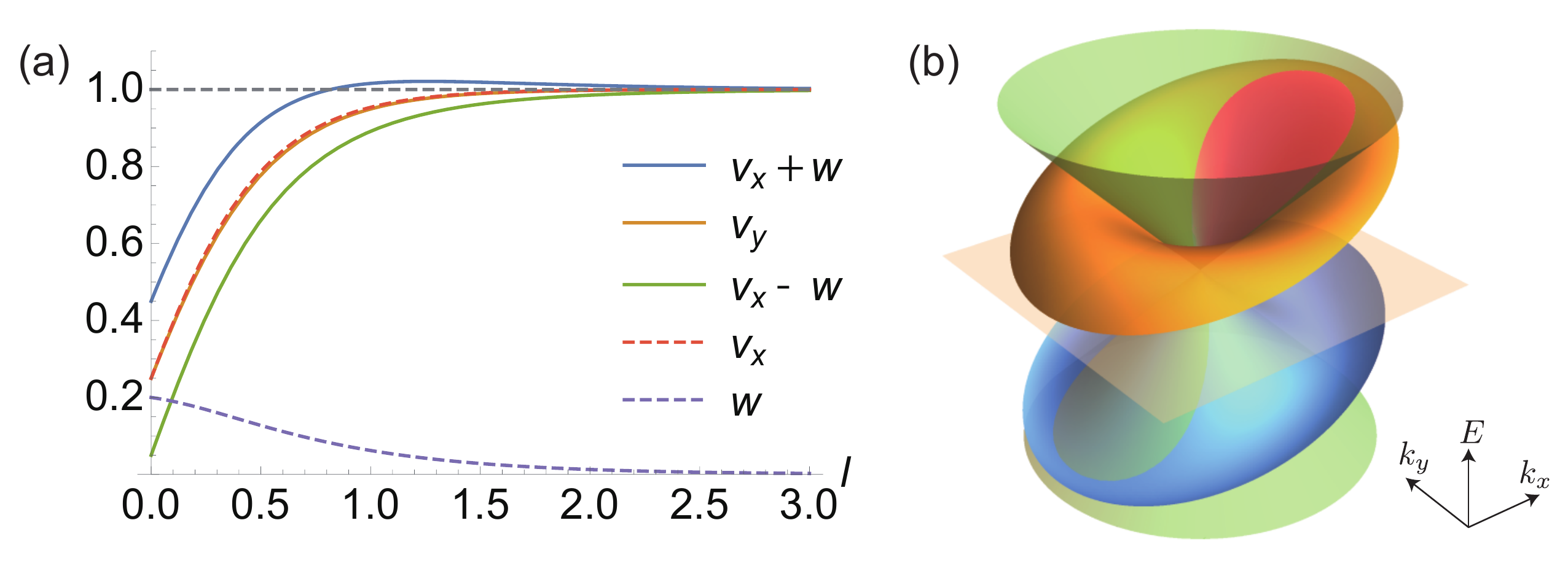}
\caption{
(a) Scale dependence of the velocities as functions of $l=(g^2/2\pi) \ln (\Lambda/\kappa)$. 
The initial values at $l = 0$ ($\kappa=\Lambda$) are given by $v_{x0} = v_{y0} = 0.25$ 
and $w_0=0.2$. 
(b) Schematic picture of the renormalized energy dispersion of a tilted WF of type I. 
The orange and blue shapes depict the conduction and valence bands, respectively, 
while the green cones are the light cones. 
In the red region of the conduction band and the corresponding part of the valence band, 
the electron velocity exceeds the speed of light $c$ and the Cherenkov radiation takes place.  
Electron propagation decays and the energy dispersion is ill-defined in those regions.  
}
\label{fig:3}
\end{figure}

The scale dependence of the velocity parameters for type I is presented in Fig~\ref{fig:3}(a), 
with the initial values $v_{x0} = v_{y0} = 0.25$ and $w_0=0.2$. 
Here we define $l=(g^2/2\pi) \ln (\Lambda/\kappa)$ with $\Lambda$ corresponding 
to a cutoff energy/momentum scale. 
Now we assume that the Weyl cone is tilted along the $x$ direction, 
$v_x+w$ corresponds to the steep slope of the energy dispersion 
and $v_x-w$ to the gentle one. 
The motion along the $y$ axis does not depend on the direction. 

For small $l$, $v_{x(y)}$ can be expanded with respect to $\ln (\Lambda/\kappa)$ as  
$v_{x(y)}(\kappa) \approx v_{x(y)0} - \beta_{v_{x(y)}} (v_{x0}, v_{y0}, w_0) \ln (\Lambda/\kappa)$. 
Note $\beta_{v_{x(y)}} (v_{x0}, v_{y0}, w_0) < 0$. 
For $w$ to be renormalized, the transverse part of the electromagnetic field needs to be relevant. 
Hence the tilt $w$ begins to be renormalized as $v_x$ and $v_y$ become larger. 
As $v_x$ approaches to the speed of light, $v_x + w$ exceeds the speed of light. 
Finally $v_{x(y)}$ and $w$ converge to 1 and 0, respectively. 
When $v_{x(y)}$ and $w$ are in the vicinity of their convergence values, 
the RG equations \eqref{eq:rg_vx}--\eqref{eq:rg_w} give $1-v_{x(y)} \propto
(\kappa/\Lambda)^{8g^2/(5\pi)}$ and $w \propto (\kappa/\Lambda)^{4g^2/(5\pi)}$. 
We can find a crossover where the logarithmic increase of $v_{x(y)}$ changes 
to the power-law convergence, i.e., the nonrelativistic regime changes to the relativistic one.  
The crossover momentum $\kappa_c$ is estimated from the relation 
$v_{x(y)0} - \beta_{v_{x(y)}} (v_{x0}, v_{y0}, w_0) \ln (\Lambda/\kappa) = 1$, 
which leads to $\kappa_c = \Lambda \exp [(1-v_{x(y)0})/\beta_{v_{x(y)}} (v_{x0}, v_{y0}, w_0)]$. 
For $v_{x0}, v_{y0}, w_0 \ll 1$, $\beta_{v_{x(y)}} (v_{x0}, v_{y0}, w_0) \approx -g^2/4$ gives 
$\kappa_c \approx \Lambda \exp (-4/g^2)$.

A key observation here is that $v_x+w$ exceeds the speed of light. 
When the phase velocity of a particle is larger than the speed of light in the material, 
the particle emits light and decays. 
This effect is know as the Cherenkov radiation~\cite{landau}. 
For the region where $v_x+w > 1$, electrons are no longer stable and decay 
with the width determined by the scattering rate of the Cherenkov radiation. 
Also the energy dispersion for this region is ill-defined, see Fig.~\ref{fig:3}(b).

\begin{figure}
\centering
\includegraphics[width=\hsize]{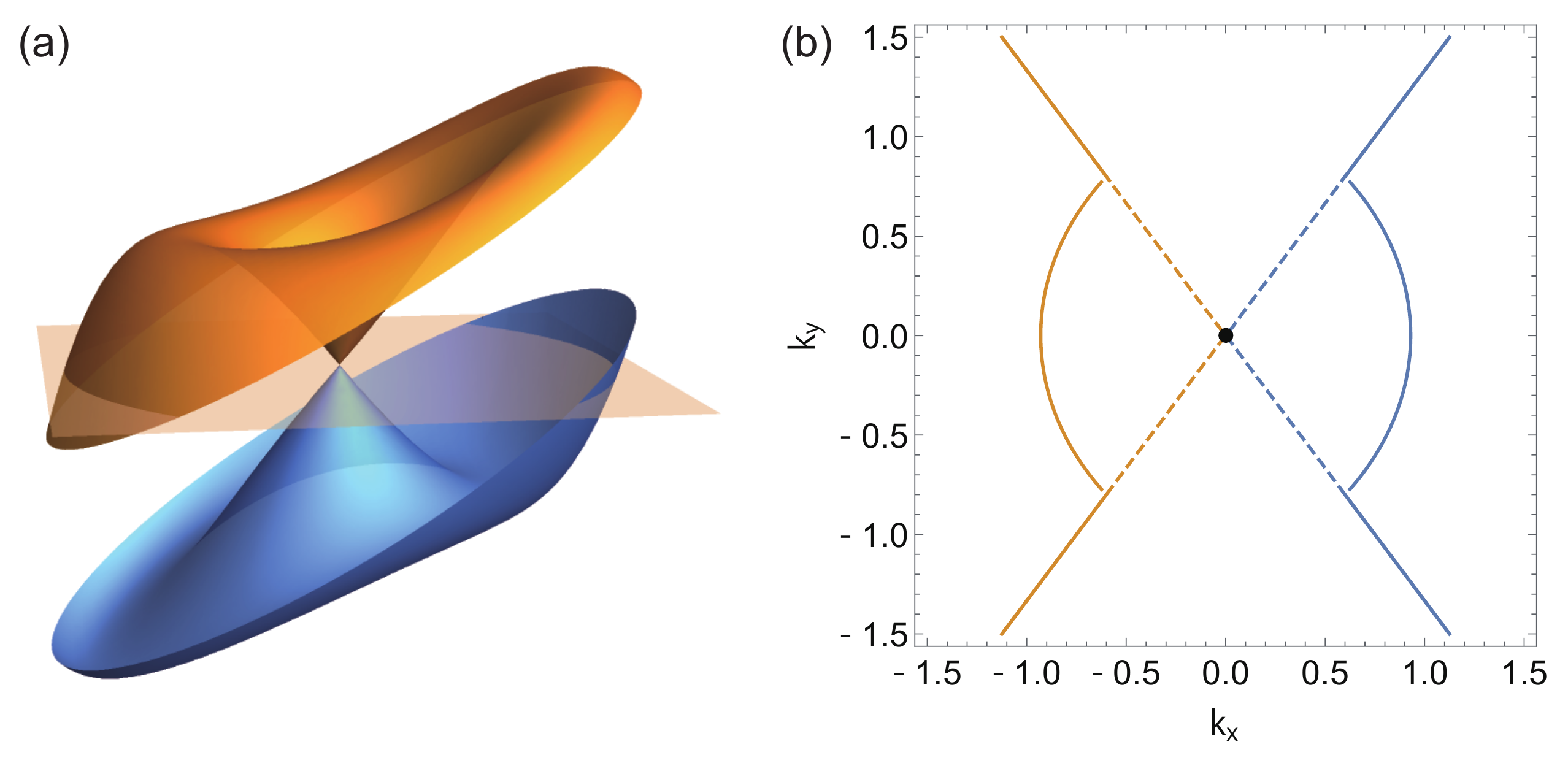}
\caption{
Effect of the electron-electron interaction for a tilted Weyl cone of type II. 
(a) Schematic picture of the energy dispersion. 
(b) Corresponding Fermi surfaces, which consist of the blue and orange curves representing 
those of the valence and conduction bands, respectively, and the black point at the Weyl point. 
}
\label{fig:4}
\end{figure}

Even for type II, the system has the same fixed point as type I. 
This means that $v_x-w$ changes its sign depending on the scale, and it accompanies 
the Lifshitz transition, namely, a change in the topology of the Fermi surface. 
The schematic energy dispersion is depicted in Fig.~\ref{fig:4}(a). 
Below certain momenta where $v_x-w=0$, the electron and hole pockets disappear, 
and the energy dispersion tends to the isotropic Dirac cone. 
The shape of the Fermi surface is schematically depicted in Fig.~\ref{fig:4}(b). 
Without the interaction effect, the electron and hole pockets are bounded by lines, and touch at $k=0$. 
However, the electron-electron interaction separates the electron and hole pockets 
at small wavenumbers. 

In the above discussion, we have not taken into account
the screening effect. Even though we consider the single
layer two-dimensional systems, the screening effect is
not negligible for type II case, where the DOS 
at the Fermi energy is finite. This gives the inverse
of the screening length $\kappa_s$ as the cutoff of
the RG. It can be estimated as
$\kappa_s a \sim e^2/(\varepsilon a W)$ with
$a$ being the lattice constant and $W$ the band width,
which can be compared with
$\kappa_s a \sim \sqrt{e^2/(\varepsilon a W)} $
for three-dimensional case.
Therefore, as long as $e^2/(\varepsilon a) \ll W$,
it is possible that $\kappa_s$ is much smaller than
$\kappa$ at which the sign change of the velocity occurs
and the Fermi surface in Fig.~\ref{fig:4}(b) is realized.

Electron interaction effects speed up the velocities at low energies, and hence modify the energy dispersion. 
The energy dispersion can be measured by angle-resolved photoemission spectroscopy. 
The DOS $D(E)$ changes as well, and it can be observed, for example, by the local 
magnetic susceptibility $\chi_s$, which is measured by nuclear magnetic resonance. 
When the Hamiltonian is spin independent, the spin susceptibility at temperature $T$ is 
obtained as~\cite{kobayashi2009}
\begin{equation}
\chi_{s} (T) 
= \int_{-\infty}^\infty dE D(E) \left( -\frac{df(E)}{dE} \right), 
\end{equation}
where $f(E)$ is the Fermi distribution. 
When electron-electron interaction is absent, the DOS $D(E)$ is proportional to the energy $E$, 
and thus the spin susceptibility is linear in temperature, $\chi_s \propto T$. 
For the tilted WFs of type I, the increase of the velocities due to the electron-electron 
interaction reduces the DOS at low energies, and hence the spin susceptibility $\chi_s$ 
is suppressed by electron interactions at low temperatures~\cite{jpsj}. 
On the other hand, for type II, there are Van Hove singularities corresponding to 
maximum, minimum and saddle points of the energy dispersion [Fig.~\ref{fig:4}(a)], 
which leads to jumps and logarithmic divergences in the DOS~\cite{vanhove}. 

The Cherenkov radiation and for type II, Van Hove singularities in addition occur for $\kappa \lesssim \kappa_c$. 
When $v_{x0}, v_{y0}, w_0 \ll 1$, $\kappa_c/\Lambda$ depends solely on the relative permittivity $\varepsilon$. 
For $\alpha$-(BEDT-TTF)$_2$I$_3$ where the conditions $v_{x0}, v_{y0}, w_0 \ll 1$ are satisfied and the linear dispersion holds up to around 10~meV~\cite{kobayashi2009}, $\varepsilon \approx 10$ makes $\kappa_c$ practically zero. 
The Cherenkov radiation and Van Hove singularities are observed below $\kappa_c$ in principle although $\kappa_c$ is usually extremely small and difficult to access experimentally.

We have investigated the effect of electron-electron interaction in the 2D tilted 
WFs including both the longitudinal and transverse electromagnetic fields. 
The RG analysis revealed that the velocities of electrons are renormalized to be 
the speed of light $c$ in the material. 
The low-energy phenomenon becomes isotropic and holds Lorentz invariance, 
which is absent in the original action. 
The result can be regarded as one of the examples of Lorentz invariance as 
low-energy emergent properties~\cite{nielsen}. 
For a strongly tilted WF with negative velocities in certain directions, the recovery of Lorentz 
invariance accompanies the change in the topology of the Fermi surface.

We thank the useful discussion with M. Hirata, K. Kanoda, H. Fukuyama, and B.-J. Yang for useful discussion. 
H. I. was supported by a Grant-in-Aid for JSPS Fellows.
This work was supported by Grant-in-Aid for Scientific Research
(No. 24224009 and No. 26103006) from the Ministry
of Education, Culture, Sports, Science and Technology
(MEXT) of Japan and from Japan Society for the Promotion of Science.


\onecolumngrid
\noindent
\hrulefill

\begin{center}
{\large\bf Supplemental Material}
\end{center}

\setcounter{equation}{0}
\setcounter{figure}{0}
\def\theequation{S\arabic{equation}}
\def\thefigure{S\arabic{figure}}

We consider the following action
\begin{equation}
S = \int d^{d_e}x \mathcal{L}_e + \int d^{d_\gamma} \mathcal{L}_\gamma ,
\end{equation}
where the electromagnetic field propagates in $d_\gamma$-D spacetime and the electron field is confined in $d_e$-D spacetime  $(d_e \leq d_\gamma)$~\cite{teber,gorbar}. 
In the present case, we set $d_\gamma = 3 + 1$ and $d_e = 2 + 1$. 
The Lagrangians for the electromagnetic field $\mathcal{L}_\gamma$, and the electron field and its coupling to the electromagnetic field are 
\begin{gather}
\mathcal{L}_\gamma = -\frac{1}{4} F^{\mu_\gamma \nu_\gamma} F_{\mu_\gamma \nu_\gamma} -\frac{1}{2a} (\partial_{\mu_\gamma} A^{\mu_\gamma})^2, \\
\mathcal{L}_e = \bar{\psi}(x) i[ \gamma^0 (D_0 + wD_1) + v_x \gamma^1 D_1 + v_y \gamma^2 D_2] \psi (x). 
\end{gather}
$D_{\mu_e}$ is the gauge covariant derivative, given by $D_{\mu_e} = \partial_{\mu_e} + ieA_{\mu_e}$. 
The indices $\mu_\gamma = 1,...,d_\gamma$ and $\mu_e = 1,...,d_e$ are used for the electromagnetic field and electron field, respectively. 
The metric tensor $g^{\mu \nu}$ is 
\begin{equation}
g^{\mu\nu} = 
\begin{pmatrix}
1 & 0 & 0 & 0 \\
0 & -1 & 0 & 0 \\
0 & 0 & -1 & 0 \\
0 & 0 & 0 & -1
\end{pmatrix}.
\end{equation}
The gamma matrix $\gamma^\mu$ obeys the anticommutation relation $\{ \gamma^\mu, \gamma^\nu\} = 2g^{\mu\nu} \mathbf{1}$, where $\mathbf{1}$ is the identity matrix.

The free fermion propagator $S_0(k)$ is defined in $d_e$ dimensions as
\begin{equation}
S_0 (k) = \frac{i}{\gamma^0(k_0 + wk_1) + v_x \gamma^1 k_1 + v_y \gamma^2 k_2 + i0^+}, 
\end{equation}
and the gauge field propagator $D_0(q_\gamma)$ in $d_\gamma$ dimensions is 
\begin{align}
D_0^{\mu_\gamma \nu_\gamma} (q_\gamma) = \frac{-ic^2}{\varepsilon (q_\gamma^2 + i0^+)} \left[ g^{\mu_\gamma \nu_\gamma} - (1-a) \frac{q_\gamma^{\mu_\gamma} q_\gamma^{\nu_\gamma}}{q_\gamma^2 + i0^+} \right]. 
\end{align}
In the reduced space where the fermions live, the reduced gauge field propagator $\tilde{D}_0^{\mu_e\nu_e} (q)$ is 
\begin{align}
\tilde{D}_0^{\mu_e \nu_e} (q_e) &= \int \frac{dq^3}{2\pi} D_0^{\mu_e \nu_e} (q_\gamma) \notag \\
&=  \frac{ic}{2\varepsilon\sqrt{-q_e^2}} \left( g^{\mu_e \nu_e} - \frac{1-a}{2} \frac{q_e^{\mu_e} q_e^{\nu_e}}{q_e^2 +i0^+} \right). 
\end{align}
We choose the Feynman gauge, i.e., $a=1$. 
The vertex $\Gamma_0^{\mu_e}$ is given by 
\begin{equation}
\Gamma_0^{\mu_e} = -iel^{\mu_e}_{\nu_e} \gamma^{\nu_e}, 
\end{equation}
where the matrix $l^{\mu_e}_{\nu_e}$ is defined by
\begin{equation}
l^{\mu_e}_{\nu_e} =
\begin{pmatrix}
1 & 0 & 0 \\
w/c & v_x/c & 0 \\
0 & 0 & v_y/c
\end{pmatrix}. 
\end{equation}

In the following analysis, we focus on the reduced space of $d_e$ dimensions, and we omit the subscript ``$e$''. 
Also we set $c=1$, as mentioned in the main text. 
The self-energy at one-loop order is given by
\begin{align}
-i \Sigma (p) = (-ie)^2 \int \frac{d^d k}{(2\pi)^d} l^{\mu\nu} \gamma_\nu S_0(k) l^{\lambda\sigma} \gamma_\sigma \tilde{D}_{0,\mu\lambda} (p-k). 
\end{align}
The divergence of this integration is regularized by the dimensional regularization; the dimension $d$ is shifted to be $d = 3-\epsilon$. 
Then we obtain the self-energy
\begin{align}
&-i \Sigma(p) \notag \\
=& (-ie)^2 \int \frac{d^d k}{(2\pi)^d} l^{\mu\nu} \gamma_\nu S_0(k) l^{\lambda\sigma} \gamma_\sigma \tilde{D}_{0,\mu\lambda} (p-k) \notag \\
=& -\frac{ie^2}{2\varepsilon} \int \frac{d^d k}{(2\pi)^d} l^{\mu\nu} \gamma_\nu \frac{i [\gamma^0 (k_0 + wk_1) + v_x \gamma^1 k_1 + v_y \gamma^2 k_2]}{(k_0+wk_1)^2 - v_x^2 k_1^2 - v_y^2 k_2^2} l^{\lambda\sigma} \gamma_\sigma  g_{\mu\lambda} \frac{1}{[-(p_0 - k_0)^2 + c^2 (p_1 - k_1)^2 + c^2 (p_2 - k_2)^2 ]^{1/2}} \notag \\
=& -\frac{ig^2}{2\pi} \left[ \frac{1}{\epsilon} + O(\epsilon^0) \right] \int_0^1 dx \frac{\sqrt{1-x}}{\sqrt{f_1(x) f_2(x)}} \notag \\
&\times \left\{ \frac{1}{f_1(x)} [ (1-x +xv_x^2 -xw^2) (1-w^2-v_x^2 -v_y^2) + xw^2 (1-w^2+v_x^2 -v_y^2) ] (\gamma^0 p_0) \right. \notag \\
&\quad -\frac{1}{f_1(x)} [ x(1-w^2-v_x^2 -v_y^2) - (1-w^2+v_x^2 -v_y^2) ] (\gamma^0 w p_1) \notag \\
&\quad -\frac{1}{f_1(x)} [ 2 v_x w (1-x +xv_x^2 -xw^2) + x v_x w (1-w^2-v_x^2 +v_y^2) ] (\gamma^1 p_0) \notag \\
&\quad -\frac{1}{f_1(x)} [ 2x w^2 - (1-w^2-v_x^2 +v_y^2) ] (\gamma^1 v_x p_1) \notag \\
&\quad \left. -\frac{1}{f_2(x)} (1-w^2+v_x^2-v_y^2) (\gamma^2 v_y p_2)
-\frac{2}{f_2(x)} w v_x v_y (\gamma^0 \gamma^1 \gamma^2 v_y p_2) \right\}, 
\end{align}
where we define $g^2 = e^2/(4\pi\varepsilon)$, and
\begin{gather}
f_1 (x) = 1- x(1-v_x^2) -x (1-x) w^2, \\
f_2 (x) = 1 -x (1-v_y^2) . 
\end{gather}

The one-loop self-energy has $\gamma^0 w p_1$ and $\gamma^0 \gamma^1 \gamma^2 v_y p_2$, which are not present in the original Lagrangian. 
When we derive RG equations, those terms will be neglected since they have only $O(g^2)$ contributions. 
From the self-energy, the following coupled RG equations are obtained: 
\begin{align}
\kappa \frac{dv_x}{d\kappa} &= -\frac{g^2}{2\pi} v_x 
	\left[ 2(1-v_x^2) F_1^0 (v_x, v_y, w) -(1-v_x^2) (1-w^2-v_x^2-v_y^2) F_1^1 (v_x, v_y, w) \right] \\
\kappa \frac{dv_y}{d\kappa} &= -\frac{g^2}{2\pi} v_y
	\left[ 2 (1-w^2-v_y^2) F_2^0 (v_x, v_y, w) -g_1(v_x, v_y, w) F_2^1 (v_x, v_y, w) +g_2(v_x, v_y, w) F_2^2 (v_x, v_y, w) \right], \\
\kappa \frac{dw}{d\kappa} &= -\frac{g^2}{2\pi} w 
	\left[ -2v_x^2 F_1^0 (v_x, v_y, w) + v_x^2 (1+ w^2-v_x^2-v_y^2) F_1^1 (v_x, v_y, w) \right], 
\end{align}
with 
\begin{gather}
F_1^n (v_x, v_y, w) = \int_0^1 dx \frac{x^n \sqrt{1-x}}{f_1^{3/2} (x; v_x, v_y) f_2^{1/2} (x; v_y)}, \\
F_2^n (v_x, v_y, w) = \int_0^1 dx \frac{x^n \sqrt{1-x}}{f_1^{3/2} (x; v_x, v_y) f_2^{3/2} (x; v_y)}, \\
g_1 (v_x, v_y, w) = (1-w^2+v_x^2-v_y^2)(1-v_x^2+w^2) + (1-w^2-v_x^2-v_y^2)(1-v_y^2) 
+ (1-v_x^2)(1-w^2-v_x^2-v_y^2) -2w^2 v_x^2, \\
g_2 (v_x, v_y, w) = (1-v_x^2) (1-v_y^2) (1-w^2-v_x^2-v_y^2) +w^2 (1-w^2+v_x^2-v_y^2) -2 w^2 v_x^2 (1-v_y^2). 
\end{gather}

\section{Analytic solutions to beta functions}

\noindent
1. $w=0$ 

For $w=0$, the RG equations become
\begin{align}
\kappa \frac{dv_x}{d\kappa} &= \frac{g^2}{2\pi} v_x (1-v_x^2) \int_0^1 dx \frac{[2-x(1-v_x^2-v_y^2)]\sqrt{1-x}}{[1-x(1-v_x^2)]^{3/2} [1-x(1-v_y^2)]^{1/2}} , \\
\kappa \frac{dv_y}{d\kappa} &= \frac{g^2}{2\pi} v_y (1-v_y^2) \int_0^1 dx \frac{[2-x(1-v_x^2-v_y^2)]\sqrt{1-x}}{[1-x(1-v_x^2)]^{1/2} [1-x(1-v_y^2)]^{3/2}}, \\
\kappa \frac{dw}{d\kappa} &= 0. 
\end{align}
We can confirm that the RG equations are symmetric under the exchange of $v_x$ and $v_y$, and the tilt stays $w=0$. 

\noindent
(a) $v_x=v_y=v$, $w=0$
\begin{align}
\kappa \frac{dv}{d\kappa} &=-\frac{g^2}{2\pi} v (1-v^2) \int_0^1 dx \frac{[2-x(1-2v^2)]\sqrt{1-x}}{[1-x(1-v^2)]^2} \notag \\
&= -\frac{g^2}{2\pi} v \left[ \int_0^1 dx \frac{\sqrt{1-x}}{(1-x+xv^2)^2} + (1-2v^2) \int_0^1 dx \frac{\sqrt{1-x}}{1-x+xv^2} \right] \notag \\
&= -\frac{g^2}{2\pi} \left[ \frac{1-4v^2}{1-v^2} +\frac{1}{v} \frac{1-2v^2+4v^4}{(1-v^2)^{3/2}}\arccos v \right]. 
\end{align}
This is consistent with the isotropic case like graphene~\cite{gonzalez}. 

\noindent
(b) $v_y=0$ and $w=0$
\begin{align}
\kappa \frac{dv_x}{d\kappa} &= \frac{g^2}{2\pi} v_x (1-v_x^2) \int_0^1 dx \frac{2-x(1-v_x^2)}{[1-x(1-v_x^2)]^{3/2}} 
= -\frac{g^2}{2\pi} v_x (1-v_x^2) \frac{2}{|v_x|}. 
\end{align}

\begin{figure}[h!]
\centering
\includegraphics[width=0.24\hsize]{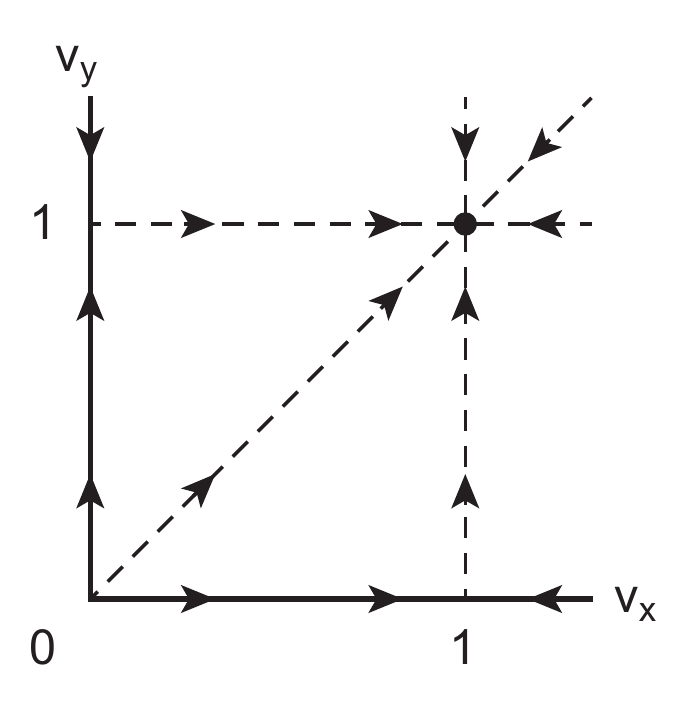}
\caption{%
RG flow obtained from the analytic solutions for $w=0$.
}
\label{fig:flow}
\end{figure}

\noindent
(c) $v_y=1$, $w=0$ 
\begin{align}
\kappa \frac{dv_x}{d\kappa} &= -\frac{g^2}{2\pi} v_x (1-v_x^2) I(v_x), 
\end{align}
where the function $I(u)$ is defined as 
\begin{align}
I(u) &= \int_0^1 dx \frac{(2+xu^2)\sqrt{1-x}}{[1-x(1-u^2)]^{3/2}} 
= 
\begin{cases}
-\dfrac{4-u^2}{(1-u^2)^2} +\dfrac{4-2u^2+u^4}{(1-u^2)^{5/2}}\text{arctanh}\left(\sqrt{1-u^2}\right) & 0<u<1, \\
\dfrac{u^2-4}{(u^2-1)^2} +\dfrac{4-2u^2+u^4}{(u^2-1)^{5/2}} \arctan \left( \sqrt{u^2-1} \right) & u>1 .
\end{cases}
\end{align}
Recall that $v_x$ and $v_y$ are symmetric when $w=0$. 
Using the results of (a)--(c), we obtain the RG flow for $w=0$ (Fig.~\ref{fig:flow}).

\noindent
2. $w\neq 0$

For $w \neq 0$, the RG equations can be analytically solved when $v_x = 1$, $v_y = 0$: 
\begin{align}
\kappa\frac{dv_x}{d\kappa} &= \kappa \frac{dv_y}{d\kappa} = 0, \\
\kappa \frac{dw}{d\kappa} &= -\frac{g^2}{2\pi} w \int_0^1 \frac{-2+w^2 x}{[1-x(1-x)w^2]^{3/2}} = \frac{g^2}{\pi} w. 
\end{align}

\end{document}